\documentstyle[11pt,aaspp4]{article}
\def\gsim{\;\rlap{\lower 2.5pt
 \hbox{$\sim$}}\raise 1.5pt\hbox{$>$}\;}
\def\lsim{\;\rlap{\lower 2.5pt
   \hbox{$\sim$}}\raise 1.5pt\hbox{$<$}\;}
\newcommand\beq{\begin{equation}}
\newcommand\eeq{\end{equation}}
\def\lya{Ly$\alpha$~}

\begin{document}

\title{Signatures of Intergalactic Dust From the First Supernovae}
\author{Abraham Loeb and Zolt\'an Haiman}
\medskip
\affil{Astronomy Department, Harvard University, 60 Garden Street,
Cambridge, MA 02138}

\begin{abstract}
We quantify the consequences of intergalactic dust produced by the first
Type II supernovae in the universe. The fraction of gas converted into
stars is calibrated based on the observed C/H ratio in the intergalactic
medium at $z=3$, assuming a Scalo mass function for the stars.  The
associated dust absorbs starlight energy and emits it at longer
wavelengths.  For a uniform mix of metals and dust with the intergalactic
gas, we find that the dust distorts the microwave background spectrum by a
$y$--parameter in the range $(0.02$--$2)\times 10^{-5} (M_{\rm SN}/0.3{\rm
M_\odot})$, where $M_{\rm SN}$ is the average mass of dust produced per
supernova.  The opacity of intergalactic dust to infrared sources at
redshifts $z\ga 10$ is significant, $\tau_{\rm dust}= (0.1$--$1)\times
(M_{\rm SN}/0.3{\rm M_\odot})$, and could be detected with the Next
Generation Space Telescope.  Although dust suppresses the Ly$\alpha$
emission from early sources, the redshifts of star clusters at $z=10$--35
can be easily inferred from the Lyman-limit break in their infrared
spectrum between 1--3.5$\mu$m.

\end{abstract}


\centerline{submitted to {\it ApJ Letters}, April 1997}

\section{Introduction}

Recent spectroscopic observations of the \lya forest at $z\ga 3$ show
evidence for a metallicity $\sim 1\%Z_\odot$ in absorbers with HI column
densities as low as $10^{15}~{\rm cm^{-2}}$ (Cowie~1996;
Songaila~\&~Cowie~1996; Tytler~et~al.~1995). Numerical simulations identify
such absorbers with mildly overdense regions in the intergalactic medium,
out of which nonlinear objects such as galaxies condense
(Hellsten~et~al.~1997, and references therein).  Indeed, damped \lya
absorbers, which are thought to be the progenitors of present--day
galaxies, show similar metalicities during the early phase of their
formation at $3.5\la z\la 4.5$ (Lu, Sargent, \& Barlow 1996).  The
universality of this mean metal abundance in absorbers spanning a range of
six orders of magnitude in HI column density, indicates that an early phase
of metal enrichment occurred throughout the universe at $z\ga 5$, before
galaxies were assembled.

The formation of an early population of star clusters at redshifts
$\sim 10$--$30$ is a natural consequence of hierarchical structure
formation in Cold Dark Matter (CDM) cosmologies (see, e.g. Haiman \&
Loeb 1997, hereafter HL97; Ostriker \& Gnedin 1996, and references
therein). The first star clusters that form in high abundance have
baryonic masses $\sim$$10^{8}M_\odot$ and virial temperatures
$\sim$$10^4$~K, which allow their gas to cool via atomic
transitions. Less massive objects with lower virial temperatures are
unable to cool and fragment into stars due to the rapid
photo-dissociation of their molecular hydrogen (Haiman, Rees, \& Loeb
1996).  Since the potential wells of the first clusters are relatively
shallow ($\sim$$10~{\rm km~s^{-1}}$), supernova--driven winds might
have expelled the metal--rich gas out of these systems and mixed it
with the intergalactic medium.  Incomplete mixing could have led to
the observed order-of-magnitude scatter in the C/H ratio along
lines-of-sight to different quasars (Rauch, Haehnelt, \& Steinmetz
1996; Hellsten~et~al.~1997). It is an interesting coincidence that the
supernova energy output associated with a metal enrichment of $\sim
1\%Z_\odot$ corresponds to $\sim 10$ eV per hydrogen atom, which is
just above the binding energy of these early star clusters. Supernova
feedback in these objects could have therefore determined the average
metallicity observed in the \lya forest. Direct observations of these
supernovae might be feasible in the future (Miralda-Escud\'e \& Rees
1997).

The measured C/H ratio can be used to calibrate the net fraction of gas
which is converted into stars by a redshift $z\approx3$.  In HL97, we
have used this fraction and an extension of the Press--Schechter formalism
to calculate the early star formation history in a variety of CDM
cosmologies.  For a large range of models we have found that the universe
is reionized by a redshift $z=10$--20, and that the resulting optical depth
to electron scattering, $\sim 0.1$, is detectable with future microwave
anisotropy experiments.  In addition, deep imaging with future infrared
telescopes, such as the Space Infrared Telescope Facility (SIRTF) or the
Next Generation Space Telescope (NGST), would be able to detect
pre--galactic star clusters. In particular, NGST should find $\ga 10^3$
star clusters at $z>10$ within its field of view of $4^\prime\times
4^\prime$, given its planned detection threshold of 1 nJy in the wavelength
range of 1--3.5~$\mu$m (Mather \& Stockman 1996).

The early epoch of star formation and metal enrichment is inevitably
accompanied by the formation of dust in supernova shells. This dust has two
important observational signatures. First, the absorption of starlight
energy and its re--emission at long wavelengths distorts the spectrum of
the cosmic microwave background (CMB) radiation (Wright et al. 1994; Bond,
Carr \& Hogan 1991; Adams et al. 1989; Wright 1981).  Second, the opacity
of the intergalactic medium to infrared sources at redshifts $z\ga 10$
could be significant.  For these redshifts, infrared in the observer frame
corresponds to UV in the source frame--a spectral regime in which dust
absorption peaks.\footnote{The effect of intergalactic dust at much lower
redshifts was discussed in the context of the reddening (Cheney \&
Rowan--Robinson 1981; Wright 1981, 1990) and the number count (Ostriker \&
Heisler 1984; Ostriker, Vogeley, \& York 1990) of quasars.} Dust
obscuration must therefore be considered when predicting the performance of
future infrared telescopes such as NGST.

In this {\it Letter}, we quantify the magnitude of the above signatures of
intergalactic dust.  In \S~2 we describe our method of calculation and in
\S~3 we compute the distortion of the microwave background spectrum and
compare it to the current COBE limit.  In \S~4 we evaluate the opacity of
the intergalactic medium to infrared sources.  Finally, \S~5 summarizes the
implications of this work for future microwave-background and infrared
observations.  Throughout the paper, we adopt a Hubble constant
$H_0=50~{\rm km~sec^{-1}~Mpc^{-1}}$ and a cosmological density parameter
$\Omega=1$.

\section{Method of Calculation}

Our first goal is to calculate the distortion of the CMB spectrum due to
the production of radiation by stars and its processing through dust.  The
CMB spectrum is initially a pure black--body with a temperature ${T_{\rm
CMB}}=2.728(1+z)$ K (Fixsen et al.  1996).  We assume that stars are born
with the Galactic initial mass function (IMF) as parameterized by Scalo
(1986), and compute their time-dependent spectra using standard spectral
atlases (Kurucz~1993) and evolutionary tracks (Schaller et al. 1992).  Our
stellar modeling is described in detail in HL97. In computing the amount of
dust produced by the stars, we assume that each type II supernova yields
${\rm 0.3M_\odot}$ of dust which gets uniformly distributed in the
intergalactic medium (see discussion in \S~3), and that the dust absorption
follows the wavelength-dependent opacity of Galactic dust.

To describe the effect of dust on the CMB we follow Wright (1981), and
define the comoving number density of photons with a comoving frequency
$\nu$ as
\beq
N_{\nu}(z)\equiv\frac{4\pi}{hc(1+z)^3}J_{\nu(1+z)}(z),
\eeq
where $J_{\nu}(z)$ is the specific intensity in ${\rm
erg~cm^{-2}~s^{-1}~Hz^{-1}~sr^{-1}}$.  The evolution of $N_{\nu}(z)$ is
determined by the radiative transfer equation:
\beq
-\frac{dN_{\nu}(z)}{dz}={cdt\over dz}
\left[j_{\nu}(z)-\alpha_{\nu(1+z)}(z)N_{\nu}(z)\right],
\label{eq:2}
\eeq
where in an $\Omega=1$ universe, $(cdt/dz)=(c/H_0)(1+z)^{-5/2}$.
Here, $j_{\nu}(z)$ is the comoving emission coefficient due to stars
and dust,
\beq
j_{\nu}(z)=\frac{4\pi}{hc(1+z)^3}j^*_{\nu(1+z)}(z) + 
\frac{8\pi}{c^3}\left\{\frac{\nu^3}
{\exp[h\nu(1+z)/k_{\rm B}T_{\rm dust}]-1}\right\}\alpha_{\nu(1+z)}(z),
\label{eq:3}
\eeq
and $\alpha_{\nu}(z)$ is the dust absorption coefficient,
\beq
\alpha_{\nu}(z)=\rho_{\rm dust}(z)\kappa_{\nu},
\label{eq:4}
\eeq
for a mass density, $\rho_{\rm dust}$, in dust.  We assume that the dust is
in thermal equilibrium with the background radiation field and equate the
power it absorbs in the UV to the power it emits in the infrared. The dust
temperature $T_{\rm dust}$ as a function of redshift is then derived from
the implicit equation,
\beq
\int_0^{\infty} d\nu  \kappa_{\nu(1+z)} \left\{ N_{\nu}(z) - 
\frac{8\pi}{c^3}\frac{\nu^3}{\exp[h\nu(1+z)/k_{\rm B}T_{\rm dust}]-1} \right\}
= 0.
\label{eq:5}
\eeq	
We adopt the dust opacity $\kappa_{\nu}$ based on the mean Galactic
extinction law $A(\lambda)/A(J)$ of Mathis (1990),
\beq
\kappa_{\nu}=
8.3\times10^3
\left[\frac{A(\lambda)}{A(J)}\right]
\left(1-\frac{0.084}{0.12+(\lambda(\mu{\rm m})-0.4)^2}\right)~~~{\rm 
cm^2~g^{-1}},
\label{eq:6}
\eeq 
where $\lambda=c/\nu$ is the wavelength; the numerical coefficient
converts the extinction per hydrogen atom, with the assumed
gas-to-dust ratio of 100, to extinction per gram of dust; and the last
term is the albedo term given by Wright (1981) for converting the
extinction (absorption plus scattering) to pure absorption.

Given a specific star--formation history, stellar spectra, and
dust--yields, one can integrate numerically
equations~(\ref{eq:3})--(\ref{eq:5}) and get the dust temperature and
the net spectrum of the radiation background at $z$=0.  As in HL97, we
express the star--formation history in terms of the mass fraction of
all baryons which are assembled into collapsed objects, $F_{\rm
coll}(z)$, and the fraction of these collapsed baryons which get
incorporated into stars, $f_{\rm star}$. The density parameter of
stars is then, $\Omega_{\rm star}(z)=f_{\rm star}F_{\rm
coll}(z)\Omega_{\rm b}$, where $\Omega_{\rm b}$ is the total baryonic
density parameter ($=0.05$ in our standard model).  We adopt the
values of both $F_{\rm coll}(z)$ and $f_{\rm star}$ from HL97; $F_{\rm
coll}(z)$ is calculated in a standard CDM cosmology based on an
extension of the Press--Schechter formalism that takes into account
pressure and photo-dissociation of ${\rm H_2}$, and $f_{\rm star}\sim
4\%$ is found from the requirement that the ${\rm C/H}$ ratio be 1\%
of the solar value at $z=3$.  The stellar emission coefficient is
found at each redshift through a convolution of the time-dependent
emission from all stars born prior to that redshift
\beq 
j^*_{\nu}(z)=f_{\rm star}\Omega_{\rm b}\rho_{\rm
c0}(1+z)^3
\int_z^{\infty} dz^{\prime}\frac{dF_{\rm coll}}{dz^{\prime}}
\epsilon^*[\nu(1+z^{\prime}),t_{z,z^{\prime}}], 
\label{eq:7}
\eeq
where $\rho_{\rm c0}$ is the current critical density of the universe,
$t_{z,z^{\prime}}$ is the time interval between redshifts $z^{\prime}$
and $z$, and $\epsilon^*(\nu,t)$ is the composite physical emissivity
(in ${\rm erg~s^{-1}~Hz^{-1}~M_{\odot}^{-1}}$) of a cluster of stars
with the Scalo IMF at time $t$ after its initial starburst (see HL97
for details).  We have assumed that all stellar photons escape their
parent clouds and are ejected into the IGM.  In reality, a fraction of
the ionizing photons will get converted into line photons with energies
$E<13.6$eV by case B recombinations inside the clouds, but this will only have
a minor effect on the total amount of energy absorbed by dust.
Free-free emission by the ionized gas in the cluster can be neglected
relative to the stellar emission.  The mass density of dust at each
redshift is given by
\beq 
\rho_{\rm dust}(z)=f_{\rm
dust}f_{\rm star}\Omega_{\rm b} \rho_{\rm c0}(1+z)^3
\int_z^{\infty}dz^{\prime}\frac{dF_{\rm coll}} {dz^{\prime}} f_{\rm
dep}(t_{z,z^{\prime}}), 
\eeq 
where $f_{\rm dust}$ is the mass fraction of
stars which gets converted into dust, and $0\leq f_{\rm dep}(t)\leq 1$
is the fraction of the total mass in dust which is deposited after a
time $t$ following the starburst (i.e. the mass fraction of the
$\geq8{\rm M_\odot}$ stars which have completed their main sequence
lifetime).  Assuming that each supernova in a Scalo mass function
produces $0.3{\rm M_\odot}$ of dust leads to $f_{\rm dust}=0.001$.
The uncertainties in this value will be discussed in the next section.

\section{Distortion of the Microwave Background Spectrum due to Dust Emission}

To parameterize the deviation of the spectrum at $z$=0 from a $T_{\rm
CMB}$=2.728 K black--body, we compute the Compton $y$--parameter
$y_c$, defined as 
\beq y_c\equiv \frac{1}{4}\left[\frac{\int d\nu
N_{\nu}}{\int d\nu N_{0,\nu}}-1 \right], 
\eeq 
where the integrals are evaluated over the FIRAS frequency range of
60--600 GHz (Fixsen et al. 1996), and 
\beq 
N_{0,\nu}=\frac{8\pi}{c^3} \frac{\nu^3}{\exp(h\nu/k_{\rm B}T_{\rm
CMB})-1}. \label{eq:10}
\eeq 

The $y$--parameter we derive is subject to a number of uncertainties
associated with the formation of dust. Each of these uncertainties can be
parameterized by a separate factor that multiplies our result for $y_{c}$.
The first uncertainty involves the average fraction of the total stellar
mass which gets converted into dust, $f_{\rm dust}$.  The value $f_{\rm
dust}=0.001$ adopted in \S~2 is equivalent to assuming that each star with
mass above $8{\rm M_\odot}$ explodes as a supernova and produces $0.3{\rm
M_\odot}$ of dust.  This mass is consistent with the values inferred from
observations of the depletion of the Si~I line in SN 1987A
(Lucy~et~al.~1991), and from the infrared echo of SN 1979C
and SN 1980K (Dwek 1983).  Theoretical estimates with 100\% condensation
efficiency predict values as high as $1{\rm M_\odot}$ of dust per supernova
(Dwek 1988; Weaver~\&~Woosley~1980).
If the dust yield of supernovae is supplemented by a significant amount of
dust produced in stars, the value of $f_{\rm dust}$ could increase even
further.  On the other hand, a substantial fraction of the dust may be
destroyed by sputtering inside the supernova envelope, or by shock waves
from other supernovae (Dwek 1997);
the $y$--parameter scales linearly with the survival fraction of dust,
$f_{\rm surv}$.  The $y$--parameter would also be affected by a
spatial correlation between the distributions of dust and stellar
sources.  If dust is correlated with star forming regions, the
radiation flux probed by the dust could be enhanced by a factor
$f_{\rm flux}$ relative to the average flux in equation~(\ref{eq:7}).
For small values of $f_{\rm flux}$, $y_c$ scales linearly with $f_{\rm
flux}$.  The final uncertainty involves the mixing efficiency of carbon
with the intergalactic gas.  The star--formation efficiency, $f_{\rm
star}$, which is required in order to get the observed metallicity of the
\lya forest is inversely proportional to this mixing efficiency.  From
equations~(\ref{eq:2}), (\ref{eq:3}) and~(\ref{eq:7}) we get $y_{\rm c}
\propto f^2_{\rm star}$, since the radiation emitted by dust is
proportional to its opacity and the stellar radiation background -- both of
which are independently proportional to $f_{\rm star}$.  All of the above
uncertainties can be combined through the definition of an overall
efficiency parameter $\eta_{\rm y}$:
\beq
\eta_{\rm y}\equiv f_{\rm surv} f_{\rm flux}
\left(\frac{f_{\rm dust}}{0.001}\right)
\left(\frac{f_{\rm star}}{0.04}\right)^2
\eeq
so that $y_{\rm c}\propto \eta_{\rm y}$, and $\eta_{\rm y}\equiv 1$ in our
standard model.

In addition to the uncertainties about the dust and the star--formation
efficiency, we note that if an early generation of low-luminosity quasars
of the type discussed by Eisenstein \& Loeb (1995) exists, these quasars
could have produced UV radiation exceeding the stellar output.  The UV
light from these quasars would be reprocessed by dust in much the same way
as the stellar radiation is, and could substantially increase the spectral
distortion derived from the stars alone.  The possible effects of early
quasars are ignored in this {\em Letter} and will be quantified in a future
paper.

Figure~1 shows the full spectrum of the background radiation for our standard
model, whose parameters are summarized in Table~1.  The solid line depicts
the total spectrum, which is a sum over the contributions from the original
CMB, the stars, and the dust, shown separately by the dotted and dashed
lines.  The net contribution from dust is negative (absorption)
for $\lambda\lsim 0.015$ cm. The dust emission at longer
wavelengths peaks around $\lambda\approx0.1$ cm, within the COBE range.
The $y$--parameter derived from this spectrum is $0.46\times10^{-5}$, a
factor of $3$ below the upper limit of $1.5\times10^{-5}$ established by
COBE (Fixsen et al.~1996).

For reference, the bottom panel of Figure~1 shows the star--formation
history, i.e. the fraction of baryons confined in stars as a function of
redshift, for our standard model.  The same panel also shows the histories
of the dust temperature and the $y$--parameter.  The dust temperature
starts to deviate noticeably from the CMB temperature (dashed line) around
$z\approx10$, when the fraction of baryons in stars 
$\Omega_{\rm star}/\Omega_{\rm b}= f_{\rm star} F_{\rm coll}\sim 1\%$.
Note that we ignore any contribution to the distortion from redshifts
$z\leq$3. Our calculation is inadequate at low redshifts, since it ignores
the enhanced production of metals and dust inside galaxies and the
contribution of quasars to the background radiation.
The spectral distortion we obtain should therefore be regarded as a
lower bound; inclusion of the contribution from lower redshifts could only
increase its magnitude.

Table~1 shows the range of values obtained for the $y$--parameter as one
deviates from the input parameters assumed in our standard model.  In
particular, if the normalization of the standard CDM power-spectrum
$\sigma_{8h^{-1}}$ is increased, then star--formation starts earlier and
spreads over a wider range of redshifts.  Although the optical depth to
dust is increased by 70\%, the corresponding change in the $y$--parameter
is only 10\%, since the average redshift delay between the production of
radiation and dust is increased.  Tilting the power spectrum to have a lower
power--law index $n$ has the opposite effect; stars are
produced later and over a narrower range of redshifts. In this case, the
optical depth is reduced by 50\% but the $y$--parameter decreases only by
10\%.  Since we normalize the star--formation efficiency by fixing the
total carbon yield of the stars, any change to the slope of the IMF has
little effect on the net production of dust.  Adding a constant
$-0.7\leq\beta\leq 1$ to the power--law index of the IMF changes the
$y$--parameter in the range $0.17\leq (y_c/10^{-5}\eta_{\rm y})\leq
0.54$.  When the photo-dissociation of ${\rm H_2}$ is ignored (e.g., due to
the renewed formation of ${\rm H_{2}}$ on dust grains), the threshold for
fragmentation and star formation in virialized objects is lowered from
virial temperatures $T_{\rm vir}\ga 10^4{\rm K}$ (where atomic line cooling
is effective) to $T_{\rm vir}\ga 10^2{\rm K}$ (where ${\rm H_2}$ cooling
operates).  In this case, star--formation starts earlier and leads to a
slight increase in the $y$--parameter.  The input parameter that affects
$y_{\rm c}$ the most is the baryon density, $y_{\rm c}\propto\Omega_{\rm
b}^2$; high values of $\Omega_{\rm b}$ (such as $\Omega_{\rm b}\ga 0.09$,
if all other parameters are held fixed) violate the existing COBE
constraint.

\begin{table}[p]
\caption{\label{tab:models} Parameter values in our standard model and
its variants, and their effect on the expected values of the
$y$--parameter and peak opacity of the intergalactic dust to a source
at $z=\infty$. The efficiency parameters $\eta_{\rm y}$ and $\eta_{\tau}$
are defined in equations (11)~and~(13).  Models with a predicted
$y$--parameter in excess of the COBE upper limit,
$y_c<1.5\times10^{-5}$, are ruled out.  The optical depth of
intergalactic dust is significant and could affect observations with
NGST.}
\vspace{0.3cm}
\begin{center}
\begin{tabular}{|c||c|c||c|c|}
\hline
Parameter & Standard & Range Considered & Compton $y_{\rm
c}/10^{-5}\eta_{\rm y}$ & Opt. Depth $\tau_{\rm dust}/\eta_{\tau}$ \\
\hline
\hline  
$\sigma_{8h^{-1}}$   & 0.67   & 0.67--1.0  & 0.42--0.46   & 0.40--0.69 \\
\hline  
$n$                  & 1.0    & 0.8--1.0   & 0.38--0.42   & 0.21--0.40 \\
\hline
IMF tilt ($\beta$)   & 0      & --0.7--1.0 & 0.17--0.54   & 0.16--0.52 \\
\hline
${\rm H_2}$ feedback & yes    & yes/no     & 0.42--0.46   & 0.40--0.77 \\
\hline  
$\Omega_{\rm b}$     & 0.05   & 0.01--0.1  & 0.02--1.71  & 0.08--0.80 \\
\hline 
\end{tabular}
\end{center}
\end{table}

\section{Optical Depth to Infrared Sources}

Next, we evaluate the optical depth to absorption and scattering by dust
along the line of sight to a source at a redshift $z_{\rm s}>3$,
\beq
\tau_{\rm dust}(\nu)=\frac{c}{H_0}\int_{3}^{z_{\rm s}}
\frac{dz}{(1+z)^{5/2}}\alpha_{\nu(1+z)}(z).
\eeq 
The absorption coefficient $\alpha_{\nu(1+z)}(z)$ is given in
equations~(\ref{eq:4})~and~(\ref{eq:6}). Since we are now
interested in dust obscuration along a specific line of sight, both
absorption and scattering must be included, and so we omit the last term
in equation~(\ref{eq:6}).  The uncertainties discussed in \S~2
concerning the formation and
survival of dust, as well as the fraction of the
baryons converted into stars, affect the optical depth through the
parameter $\eta_{\tau}$,
\beq
\eta_{\tau}\equiv f_{\rm surv} f_{\rm flux}
\left(\frac{f_{\rm dust}}{0.001}\right)
\left(\frac{f_{\rm star}}{0.04}\right),
\eeq
so that $\tau_{\rm dust}$ obtains the value predicted by our standard model
times $\eta_{\tau}$.

Figure~2 shows the resulting optical depth in our standard model as a
function of observed wavelength $\lambda$, for three different source
redshifts.  The optical depth peaks at $\lambda\sim 1\mu$m, where it
reaches a value $\tau_{\rm dust}=0.4$.  This high value could
suppress the Ly$\alpha$ emission peak in the spectra of early star clusters
(see Charlot \& Fall 1991); attempts to detect this emission line from
galaxies at $z\sim$ 2--4 proved difficult in the past (Thompson, Djorgovski
\& Trauger 1995; Lowenthal et al. 1997).  
Since the total amount of dust is the same in all models with the same IMF,
all other variations in the peak optical depths given in Table~1 arise from
differences in the production history of dust.  In particular, the optical
depth is high when stars form early (e.g., as a result of a high
$\sigma_{8h^{-1}}$ or the lack of ${\rm H_2}$ destruction), and low when
stars form late (low $n$).  We note that the optical depths we derive are
below the values that could cause a measurable reddening in the spectra of
quasars at $z\la5$ (Rowan--Robinson 1995).

To illustrate the potential effect of dust on future observations with
NGST, we show the composite spectrum of two star--clusters, located at
$z_{\rm s}$=10 and 20, and containing $10^8~{\rm M_{\odot}}$ and
$4.6\times10^8~{\rm M_{\odot}}$ in stars, respectively (bottom panel).
The stars are assumed to be distributed on the main sequence according
to the Scalo IMF.  In both cases the top curve shows the original
unprocessed spectrum, and the bottom curve shows the obscured spectrum
due to the intervening dust in the standard model (with $\eta_{\rm
y}=1$).  These results imply that the performance of NGST near its
planned detection threshold of 1 nJy in the wavelength range
1--3.5~$\mu$m, could be affected by intergalactic dust.

Another important conclusion that can be drawn from Figure 2 is that
despite the obscuration by dust, the Lyman-limit break could easily be
identified in the emission spectrum of star clusters at high-redshifts.
Additional absorption by neutral hydrogen in the IGM would only make the
break appear sharper.  For NGST observations between 1--3.5$\mu$m, the
detection of this break can identify source redshifts in the range $z_{\rm
s}=$10--35.  The feasibility of this {\it UV dropout} technique was
recently demonstrated through the successful photometric identification of
galaxy redshifts in the interval $2.5 \la z_{\rm s}\la 3.5$ (Steidel et al.
1996; Madau et al. 1996).

\section{Summary}

We have quantified the imprint of intergalactic dust due to the first stars
on the $y$--parameter of the CMB spectrum (in the COBE wavelength range),
$y_{\rm c}$, and on the opacity to absorption and scattering of radiation
from sources at high redshifts, $\tau_{\rm dust}(z_{\rm s})$.  We
normalized the star-formation efficiency so as to get the observed
$\sim$1\% solar C/H ratio in Ly$\alpha$ absorbers at $z$=3.  With this
normalization, we find $y_{\rm c}\sim10^{-5}$ and $\tau_{\rm
dust}(\infty)\sim 1$ for plausible choices of the cosmological parameters
and stellar properties at high redshifts (see Table 1).  Uncertainties due
to the mixing efficiency of metals with the IGM, the stellar IMF, the
destruction of dust, the spatial correlations between dust and stellar
sources, or the cosmological parameters, could change these estimates by an
order of magnitude.  If early quasars exist, their additional UV radiation
would enhance this spectral distortion.

The current COBE limit, $y_{\rm c}<1.5\times 10^{-5}$, already places
interesting constraints on the dust formation history of the universe at
$z\ga 10$. Complementary microwave anisotropy limits on the scale of $\sim
1^{\prime\prime}$ would constrain the degree of patchiness in the dust
distribution.  But most importantly, the future detection of source
reddening at $z_{\rm s}\ga 10$ by NGST, would be instrumental in
establishing the properties of the first stars in the universe. Despite the
significant dust opacity, Figure 2 shows that infrared photometry by NGST
could identify the redshift of sources at $10\la z_{\rm s}\la 35$, based on
their pronounced Lyman-limit break at observed wavelengths of 1--3.5$\mu$m.

\acknowledgements

We thank Dimitar Sasselov and Eli Dwek for useful discussions. This work
was supported in part by the NASA ATP grant NAG5-3085 and the Harvard
Milton fund.


\clearpage
\newpage
\begin{figure}[b]
\vspace{2.6cm}
\includegraphics{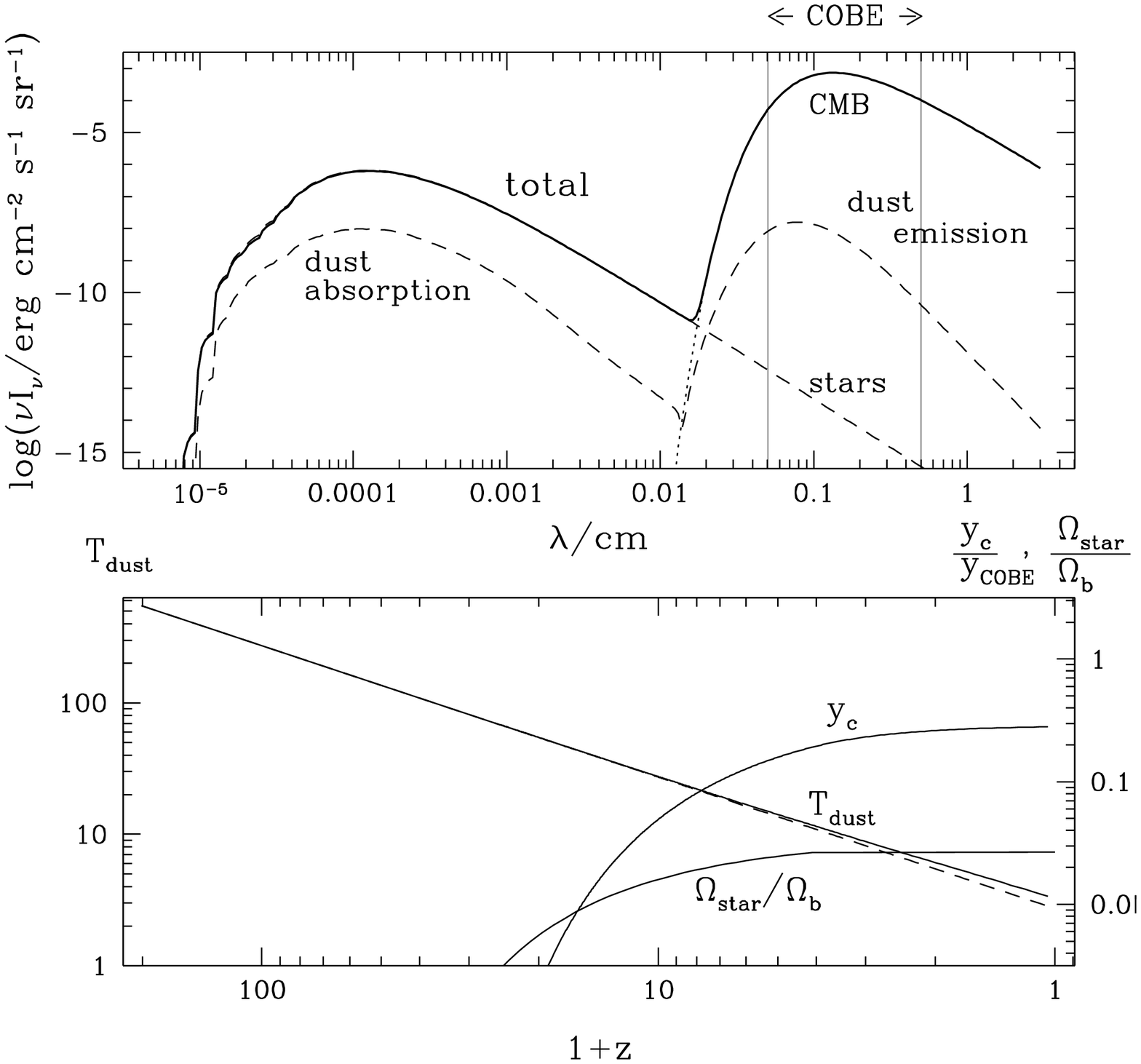}
\vspace*{4.5in}
\caption[Spectrum] {\label{fig:spectrum} {\bf Top panel}: the full
spectrum at $z$=0 in our standard model for the star--formation
history.  The solid line shows the total spectrum, the dotted and
dashed lines show the separate contributions of the original CMB, the
stars, and the dust.  The net contribution of dust results in
absorption at $\lambda\lsim 0.015$ cm, and emission at longer
wavelengths. {\bf Bottom panel}: the redshift evolution of the
fraction of baryons in stars, the dust temperature, and the
$y$--parameter.  Note that $f_{\rm star}\equiv F_{\rm
coll}(z)^{-1}\Omega_{\rm star}/\Omega_{\rm b}$.  The dashed curve
shows the history of the CMB temperature.}
\end{figure}

\clearpage
\newpage
\begin{figure}[b]
\vspace{2.6cm}
\includegraphics{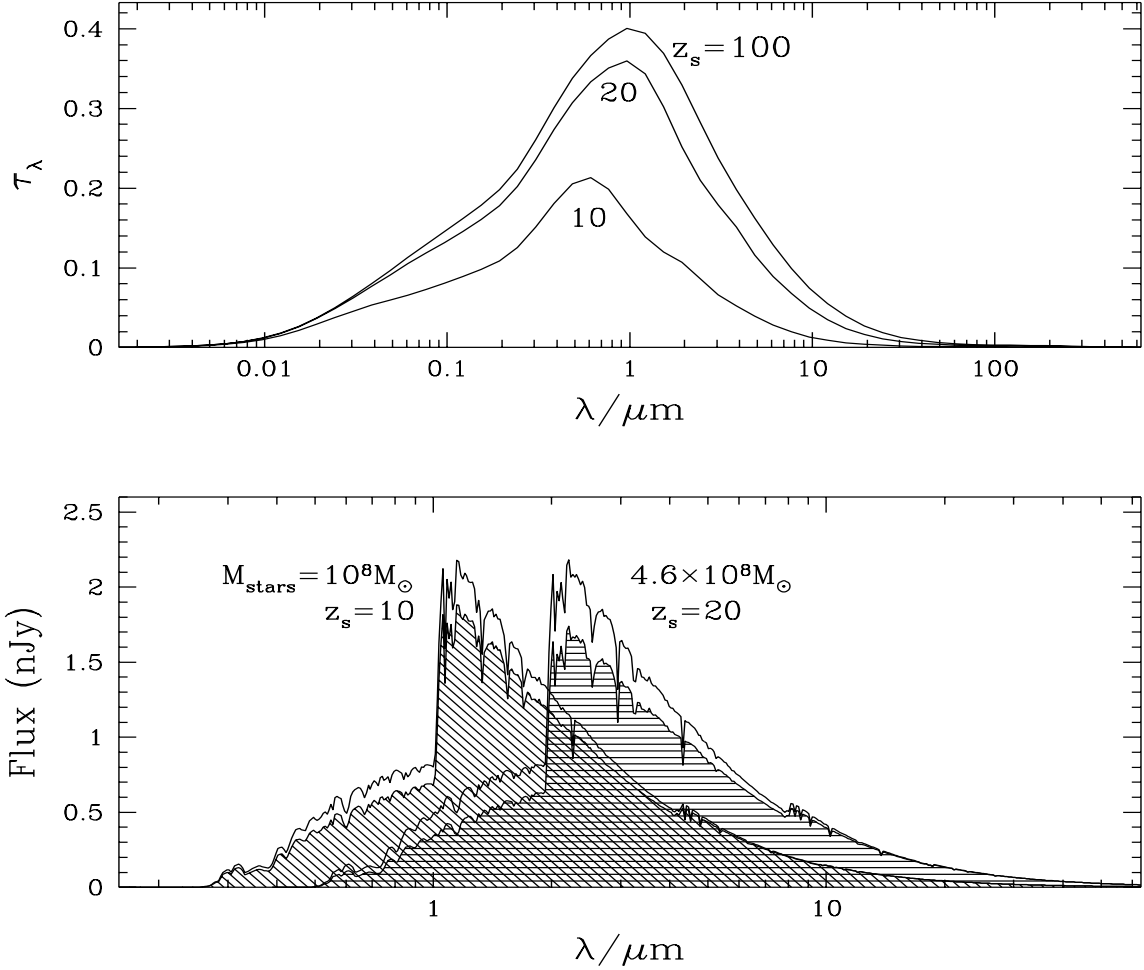}
\vspace*{4.5in}
\caption[Tau] {\label{fig:tau} {\bf Top panel}: optical depth in our
standard model to absorption and scattering by dust at $z\geq3$, as a
function of observed wavelength.  The source is located at redshifts
$z_{\rm s}$=$10$, $20$, or $100$.  {\bf Bottom panel}: the expected flux
with (lower curve) and without (upper curve) intervening dust, from
star--clusters at $z_{\rm s}$=10 and $20$, containing $10^8~{\rm
M_{\odot}}$ and $4.6\times10^8~{\rm M_{\odot}}$ in stars. The stars are
assumed to be distributed on the main sequence according to a Scalo IMF.
The break in the spectra at the redshifted Lyman limit wavelength
includes only the absorption by neutral hydrogen in the stellar
atmospheres.}
\end{figure}

\end{document}